\documentclass{pnastwo-3}

\usepackage{graphicx}
\usepackage{times}
\usepackage{amssymb,amsfonts,amsmath}

\usepackage{alphalph}

\newcommand{\apj}{Astrophys. J. }
\newcommand{\apjl}{Astrophys. J.}
\newcommand{\apjs}{Astrophys. J. Suppl. Ser. }
\newcommand{\mnras}{Monthly Not. Royal Astron. Soc. }
\newcommand{\solphys}{Solar Phys. }

\newcommand{\aap}{Astron. Astrophys. }
\newcommand{\nat}{Nature }
\newcommand{\aj}{Astron. J. }
\newcommand{\kms}{km\,s$^{-1}$}

\contributor{Submitted to Proceedings of the National Academy of Sciences of the United States of America}
\url{www.pnas.org/cgi/doi/}
\copyrightyear{2012}
\issuedate{Issue Date}
\volume{Volume}
\issuenumber{Issue Number}
\begin{document}

\title{Seismic constraints on rotation of sun-like star and mass of exoplanet}

\author{
Laurent Gizon\affil{1}{Max-Planck-Institut f\"ur Sonnensystemforschung, 37191 Katlenburg-Lindau, Germany}\affil{2}{Institut f\"ur Astrophysik, Universit\"at G\"ottingen, 37077 G\"ottingen, Germany},
J\'erome Ballot\affil{3}{Institut de recherche en astrophysique et plan\'etologie, CNRS, 31400 Toulouse, France}
\affil{4}{Universit\'e de Toulouse, UPS-OMP, 31400 Toulouse, France},
 Eric Michel\affil{5}{LESIA, Observatoire de Paris, CNRS UMR8109, Universit\'e Pierre et Marie Curie, Universit\'e Denis Diderot, 92195 Meudon, France},
 Thorsten Stahn\affil{2}{}\affil{1}{},
G\'erard Vauclair\affil{3}{}\affil{4}{},
Hans Bruntt\affil{5}{},
Pierre-Olivier Quirion\affil{6}{Agence Spatiale Canadienne, Saint-Hubert, J3Y 8Y9 Qu\'ebec, Canada},
Othman~Benomar\affil{7}{Sydney Institute for Astronomy (SIfA), School of Physics, University of Sydney, NSW 2006, Australia}
\affil{8}{Institut d'Astrophysique Spatiale, Universit\'e Paris Sud\,--\,CNRS (UMR8617), Batiment 121, 91405 ORSAY Cedex, France},
Sylvie Vauclair\affil{3}{}
\affil{4}{},
Thierry Appourchaux\affil{8}{},
Michel Auvergne\affil{5}{},
Annie Baglin\affil{5}{},
Caroline Barban\affil{5}{},
Fr\'ederic Baudin\affil{8}{},
Micha\"el Bazot\affil{9}{Centro de Astrof\'isica da Universidade do Porto, 4150-762 Porto, Portugal},
Tiago~L. Campante\affil{9}{}
\affil{10}{Danish AsteroSeismology Centre, Department of Physics and Astronomy, University of Aarhus, 8000 Aarhus C, Denmark}
\affil{11}{School of Physics and Astronomy, University of Birmingham, Edgbaston, Birmingham B15 2TT, UK},
Claude Catala\affil{5}{},
William~J. Chaplin\affil{11}{},
Orlagh~L. Creevey\affil{12}{Universidad de La Laguna, Departamento de Astrof{\'i}sica, 38206 La Laguna, Tenerife, Spain}
\affil{13}{Instituto de Astrof{\'i}sica de Canarias, 38205 La Laguna, Tenerife, Spain}
\affil{14}{Laboratoire Lagrange, UMR7293, Universit\'e de Nice Sophia-Antipolis, CNRS, Observatoire de la C\^ote d'Azur, Nice, France},
S\'ebastien Deheuvels\affil{3}{}\affil{4}{},
No\"el Dolez\affil{3}{}\affil{4}{},
 Yvonne Elsworth\affil{11}{},
 Rafael~A. Garc\'ia\affil{15}{Laboratoire AIM, IRFU/SAp\,--\,\-CEA/DSM\,--\,\-CNRS\,--\,\-Universit\'e Paris Diderot, 91191 Gif-sur-Yvette Cedex, France},
 Patrick Gaulme\affil{16}{Department of Astronomy, New Mexico State University, P.O. Box 30001, MSC 4500, Las Cruces, NM 88003-8001, USA},
 St\'ephane Mathis\affil{15}{},
 Savita Mathur\affil{17}{High Altitude Observatory, NCAR, P.O. Box 3000, Boulder, CO 80307, USA},
 Beno\^it Mosser\affil{5}{},
 Clara R\'egulo\affil{12}{}\affil{13}{},
Ian~W. Roxburgh\affil{18}{Queen Mary University of London, Astronomy Unit, School of Physics and Astronomy, London E1~4NS, UK},
David Salabert\affil{14}{},
 R\'eza Samadi\affil{5}{},
Kumiko~H. Sato\affil{15}{},
Graham~A. Verner\affil{11}{}\affil{18}{},
Shravan~M. Hanasoge\affil{19}{Department of Geosciences, Princeton University, NJ 08544, USA}\affil{1}{}
\and
Katepalli~R. Sreenivasan\affil{20}{Physics Department and Courant Institute of Mathematical Sciences, New York University, NY 10012, USA}
%\affil{22}{To whom correspondence should be addressed; E-mail: gizon@mps.mpg.de and sv@nyu.edu.}
}

\contributor{Submitted to Proceedings of the National Academy of Sciences
of the United States of America}

%%%%%%%%%%%%%%%%% END OF PREAMBLE %%%%%%%%%%%%%%%%
\maketitle

\begin{article}

\begin{abstract}
Rotation is thought to drive cyclic magnetic activity in the Sun and Sun-like stars. Stellar dynamos, however, are poorly understood owing to the scarcity of observations of rotation and magnetic fields in stars. Here, inferences are drawn on the internal rotation of a distant Sun-like star by studying its global modes of oscillation. We report asteroseismic constraints imposed on the rotation rate and the inclination of the spin axis of the Sun-like star HD\,52265, a CoRoT prime target known to host a planetary companion. These seismic inferences are remarkably consistent with an independent spectroscopic observation (rotational line broadening) and with the observed rotation period of starspots.
%, but the seismic estimates have significantly smaller uncertainties.
Further, asteroseismology constrains the mass of exoplanet HD\,52265b. Under the standard assumption that the stellar spin axis and the axis of the planetary orbit coincide, the minimum spectroscopic mass of the planet can be converted into a true mass of $1.85^{+0.52}_{-0.42} M_{\rm Jupiter}$---which implies that it is a planet, not a brown dwarf.
\end{abstract}


\abbreviations{CoRoT, Convection Rotation and planetary Transits}

\dropcap{S}pace photometry has made possible high-precision seismology of Sun-like stars \cite{WIRE,Michel2008,Chaplin2011,Ballot2011,Appourchaux2012}. Precise measurements of the frequencies of the global modes of acoustic oscillations place tight constraints on the internal structure of these stars \cite{JCD2010}. For example, improved stellar parameters are used to refine the physics of stellar interiors and find many applications in astrophysics. Accurate determinations of the radii, masses, and ages of planet-host stars are essential for the characterization of exoplanets detected in transits \cite{Batalha2011,Kepler21b2012,Kepler22b2012}.

In this paper we present the first unambiguous measurement of mean internal rotation (rotation averaged over the entire stellar interior) and inclination angle of a Sun-like star by means of asteroseismology. Internal rotation is a fundamental physical property of stars: it affects stellar evolution through increased mixing of chemicals and mass loss, and is responsible for magnetic activity \cite{Noyes1984b}. Our study extends earlier studies of the seismic signature of rotation in $\alpha$ Centauri A \cite{Fletcher2006,Bazot2007} and in red giant stars \cite{Beck2012,Deheuvels2012}.

The star HD\,52265 was observed continuously for 117~days between November 2008 and March 2009 in the asteroseismic field of the space telescope CoRoT \cite{Baglin2006}. This relatively bright star (visual magnitude $6.3$) was selected as a primary target because it hosts a planetary companion, which was detected through the wobbling of the star via the radial velocity method \cite{Butler2000,Naef2001,Butler2006}. With an effective temperature \cite{Ballot2011} of $6100\pm 60$~K and an absolute luminosity \cite{Ballot2011} of $2.09\pm 0.24 L_{\rm Sun}$, HD\,52265 is a main-sequence G0V star. Combining these two values, the stellar radius is deduced to be $1.30 \pm 0.08 R_{\rm Sun}$. Isochrone fits \cite{Gonzalez2001,Valenti2005} give a stellar mass near $1.2$~$M_{\rm Sun}$ with a typical error of $0.05$~$M_{\rm Sun}$ and a stellar age between $2.1$ and $2.7$~Gyr. The planetary companion, HD~52265b, orbits the star with a period of 119~days, a semi-major axis of $0.5$~AU, and a minimum mass $M_{\rm p}\sin i_{\rm p} = 1.09\pm0.11 M_{\rm Jupiter}$, where $M_{\rm p}$ is the true mass of the planet and $i_{\rm p}$ the inclination of the orbital axis to the line of sight \cite{Butler2006}. HD\,52265 is overmetallic with respect to the Sun ($[{\rm M}/{\rm H}]=0.19\pm0.05$), a property that has been associated with the formation of hot Jupiters \cite{Fisher2005}. See Table 1 for a summary of basic properties.

\begin{figure}
\label{fig.power}
\centerline{
\includegraphics[angle=90,width=\linewidth]{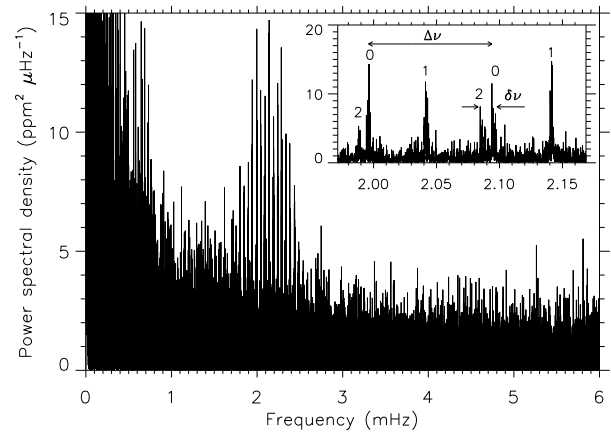}
}
\caption{Power spectrum of global acoustic oscillations of HD~52265. The p-modes have measurable power in the range $1.6$-$2.6$~mHz. At low frequencies, the power is due to stellar convection and magnetic activity; at high frequencies it is dominated by photon noise. The inset shows the power spectrum in the interval $1.97$\,--\,$2.17$~mHz, where acoustic modes are labeled with their spherical harmonic degrees, ${l}\leq 2$. Indicated are the large frequency separation between consecutive radial modes, $\Delta\nu$ (sensitive to mean density), and the small frequency separation between adjacent radial and quadrupole modes, $\delta\nu$ (sensitive to age). Mode identification is unambiguous by virtue of the analogy with the solar spectrum.}
\end{figure}

The power spectrum of the HD~52265 CoRoT data exhibits a series of peaks near $2$\,mHz caused by global acoustic oscillations (Figure~1). As in the Sun and other stars with outer convection zones, these oscillations are continuously excited by near-surface convection \cite{Michel2008}. The star oscillates in high-overtone modes whose horizontal spatial patterns are given by spherical harmonics $Y_l^m$, where $l$ is the harmonic degree and $m$ the azimuthal order with $-l\le m \le l$. The comb-like structure of the power spectrum is due to a repeating sequence of pulsations observable in quadrupole ($l=2$), radial ($l=0$), and dipole ($l=1$) modes. The $l=2$ and $l=0$ modes are remarkably well resolved in frequency space, and enable unambiguous identification of modes.

To first order \cite{Tassoul1980}, the spectrum of the mode frequencies is specified by two characteristic frequencies (inset to Figure~1): the ``large frequency" separation, $\Delta\nu$, and the ``small frequency" separation, $\delta\nu$. The large frequency separation between consecutive $l=0$ modes is the inverse of the sound travel time across a stellar diameter. The small frequency separation between adjacent $l=2$ and $l=0$ modes is sensitive to the radial gradient of sound speed in the nuclear-burning core, thus to helium content and the age of the star \cite{Gough1987}.

Figure~2a shows the power spectrum in \'echelle format \cite{Grec1983} of HD~52265 using a folding frequency of $98.5$~$\mu$Hz. Modes have measurable power over ten consecutive oscillation overtones (radial orders), $16\le n \le 25$. For comparison, Figure~2b shows the \'echelle spectrum of the Sun computed using 117~days of SoHO/VIRGO \cite{Froehlich1995} data and a folding frequency of $135.3$~$\mu$Hz. Although photon noise is higher for HD~52265, the similarity between the two power spectra is striking. HD~52265 is an object that pulsates like the Sun, though there are measurable differences.

\begin{figure}
\label{fig.echelle}
\centerline{
\includegraphics[angle=90,width=\linewidth]{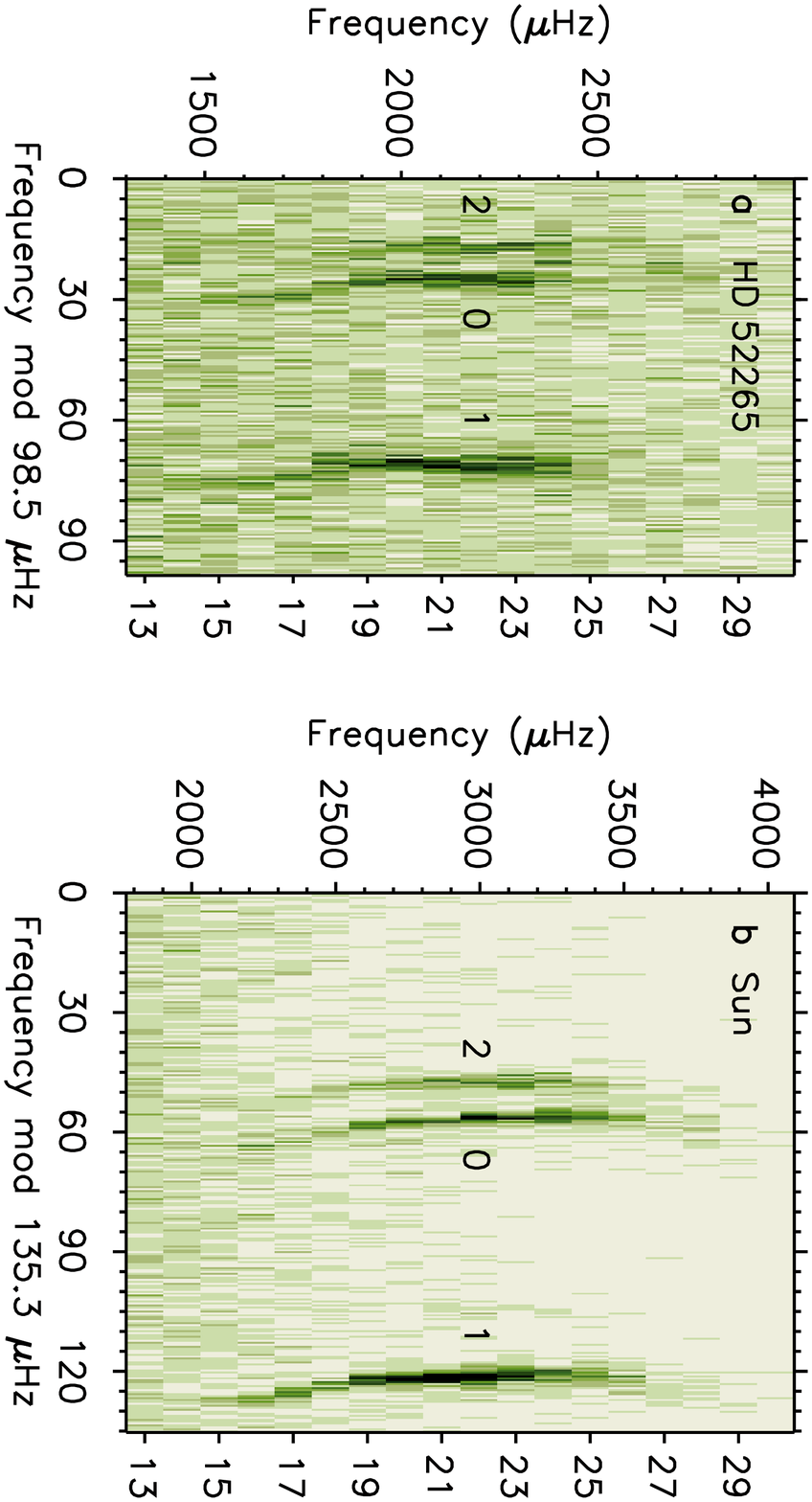}
}
\caption{\'Echelle spectrum and comparison with the Sun. (a) \'Echelle spectrum of HD 52265 using a folding frequency of $98.5$~$\mu$Hz. The power spectrum is cut into frequency segments which are stacked in the vertical direction. Integers along the right axis indicate the number of frequency segments, starting from zero frequency, i.e. the radial order of the $l=0$ modes. The nearly vertical ridges of power (labeled according to spherical harmonic degree $l$) indicate that the folding frequency is close to the large separation $\Delta \nu$. (b) For comparison and mode identification, we show the \'echelle spectrum of the Sun using 117~days of SoHO/VIRGO photometry \cite{Froehlich1995} (green channel) and a folding frequency of $135.3$~$\mu$Hz.}
\end{figure}

\section*{Results}
\subsection*{Global fit of the power spectrum.}
We estimate the parameters of the individual modes of oscillation by fitting a global parametric model to the power spectrum using a maximum-likelihood technique \cite{Anderson1990}, by considering that the probability density function of the power at any given frequency is an exponential distribution. All modes with $l\le2$ are fitted together in the range $1.6-2.6$~mHz. The expectation value of the power spectrum of each individual mode is modeled by a Lorentzian, whose height and width are allowed to vary with frequency. In order to improve the robustness of the fit, the large and small frequency separations are taken to be smooth (polynomial) functions of the radial order. The noise background is modeled as the sum of a convective component and a white-noise component. As for the Sun, the convective component is well approximated by the sum of two Lorentzians, representing the two dominant timescales of convection, granulation and supergranulation \cite{Michel2008}.

We include the effects of stellar rotation in the model. First, rotation removes the frequency degeneracy of the $(2l+1)$ azimuthal components. Assuming a slowly rotating star, the frequency of mode $(n,l,m)$ may be approximated by
\begin{equation}
\nu_{nlm}=\nu_{nl}+m\Omega/2\pi,
\end{equation}
where $\Omega$ is a suitable radial average of the angular velocity over the star \cite{Hansen1977}. Second, mode visibility depends on the angle $i$ between the rotation axis and the line of sight. Assuming (as for the Sun) energy equipartition in a multiplet $(l,n)$ between azimuthal components, mode power is proportional to
\begin{equation}
E_l^m(i) = \frac{(l-|m|)!}{(l+|m|)!} \left[ P_l^m(\cos i) \right]^2
\end{equation}
where the $P_l^m$ are associated Legendre functions \cite{Gizon2003}. For example, for dipole modes, $E_1^0(i) = \cos^2 i$ and $E_1^{\pm 1}(i)=1/2 \sin^2 i$. Thus, the ratios in mode power between azimuthal components tell us about $i$, while the splitting between mode frequencies tells us about $\Omega$.

The fitted multiplets are shown in Figure~3. The values of $\Omega$ and $\sin i$ are inferred at the same time as the mode frequencies and other mode parameters. The random errors associated with the estimated mode parameters are deduced from Monte-Carlo simulations.

\begin{figure}
\label{fig.fits}
\centerline{\includegraphics[width=\linewidth]{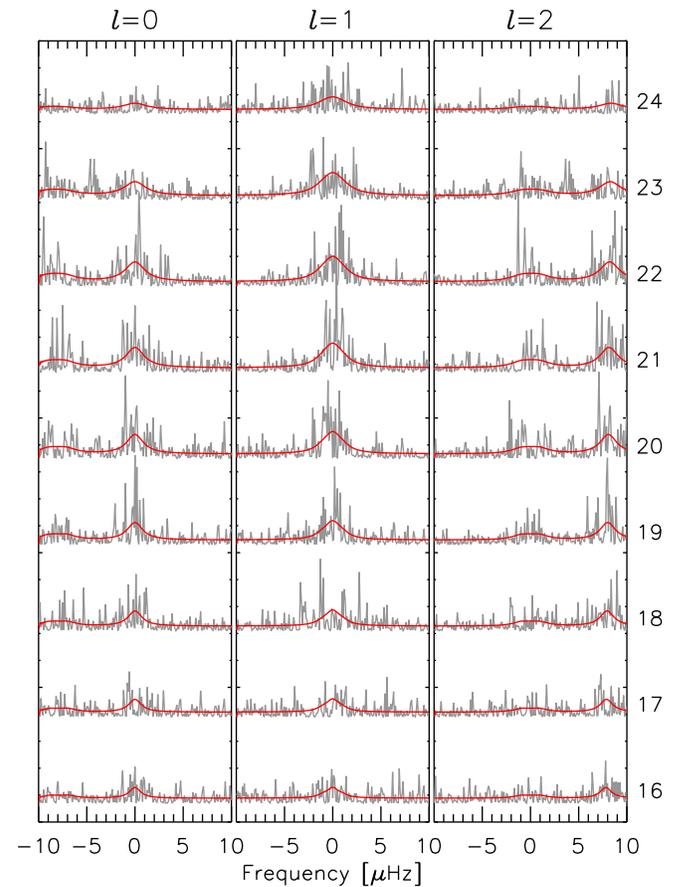}}
\caption{
Power spectra of the radial (${l}=0$, left panel), dipole (${l}=1$, middle), and quadrupole modes (${l}=2$, right) of HD 52265 (gray curves). Nine consecutive radial orders $16\leq n \leq 24$ are shown, with $n$ decreasing from top to bottom. In each panel the frequency axis is shifted by the central frequency of the multiplet, $\nu_{n l}$, obtained from the global fit. The global fit (red curves) is an estimate of the expectation value of the power spectrum and includes the effects of rotation on oscillations. Each azimuthal component ($n, l, m$) in a multiplet is modeled by a Lorentzian line profile, describing damped harmonic oscillation.}
\end{figure}

\subsection*{Seismic stellar model.}
In the frequency range $1.85$\,--\,$2.30$~mHz where the mode power is the largest, we measure an average large frequency separation of $\Delta\nu=98.56\pm 0.13$~$\mu$Hz and an average small frequency separation of $\delta\nu=8.08\pm0.16\,\mu$Hz. The large frequency separation is proportional to the square root of the mean stellar density, implying that HD~52265 is less dense than the Sun by a factor $0.5338\pm0.0014$. This seismic constraint on the mean stellar density is sixty times more precise than the spectroscopic constraint.

We have estimated the fundamental stellar properties of HD~52265 by finding the best-fit stellar model among an extended grid of stellar models computed with the Aarhus Stellar Evolution Code (ASTEC). Using the average large and small frequency separations given above together with the observed effective temperature of the star and its metallicity (Table 1), the SEEK optimization procedure \cite{Quirion2010} returns a best-fit stellar model with a seismic radius $R=1.34\pm 0.02 R_{\rm Sun}$ and a seismic mass $M=1.27\pm 0.03 M_{\rm Sun}$, where formal error bars are several-fold smaller than the classical ones (Table~2). The seismic age is $2.37 \pm 0.29$~Gyr, where we quote the formal error. By comparison, stellar ages deduced from isochrone fits lead to typical errors \cite{Saffe2005} of 30\,--\,50\%, which are significantly worse. Detailed asteroseismic modeling of HD~52265 not only places tight constraints on the mass, radius, and age of the star but also on its initial chemical composition \cite{Escobar2012}.

\subsection*{Internal stellar rotation.}
The global fit of the power spectrum returns a rotational splitting frequency $\Omega/2\pi=0.98^{+0.19}_{-0.29}$~$\mu$Hz and the inclination $\sin i=0.59^{+0.18}_{-0.14}$. The inferred rotational splitting of HD~52265 is about $2.3$ times larger than that of the Sun (see Table 2).

To further test our methodology, we average the power spectrum
over multiplets with the same $l$ value in order to smooth out random
variations of power with frequency due to the realization noise (Figure 3).
This average is computed over nine consecutive radial orders ($16 \leq n \leq 24$), after shifting multiplets by their central frequencies $\nu_{nl}$ using Fourier interpolation. A similar averaging procedure was used in the early days of helioseismology \cite{Grec1980} to measure the small frequency separation $\delta \nu$.

The average spectra of HD 52265 for $l = 0$, $l = 1$, and $l = 2$ are shown in Figure 4. The widths at half maximum of the profiles for $l=1$ and $l=2$ modes are larger than that of the $l=0$ singlet by $20$\% and $90$\%, respectively. Thus the non-radial multiplets are broadened by rotation, confirming that stellar rotation (see Table 2) has a measurable effect, as inferred rigorously from the maximum likelihood model fit. However, the individual $m$ components are not resolved due to the intrinsic line width of the modes, which (near maximum power) is about twice as large as the rotational splitting---a situation comparable to that of the Sun.

Our seismic estimates of $\Omega$ and $i$ are remarkably consistent with independent measurements from spectroscopy and starspot rotation, as shown in Figure~5. The observed \cite{Ballot2011} spectroscopic rotational velocity $v\sin i=3.6^{+0.3}_{-1.0}$\,\kms is fully consistent with the seismic value $R\,\Omega \sin i = 3.4\pm 0.8$~\kms (surface and bulk rotation rates of solar-type stars are expected to be similar, as in the Sun). In addition, the photometric time series of HD\,52265 is modulated by the rotation of starspots at two prominent periods of $10.8$ and $12.7$~days, corresponding to the cyclic frequencies $\Omega_{\rm spot}/2\pi = 0.91$ and $1.07$~$\mu$Hz, values that are consistent with seismology. The two periods may be associated with starspots at different latitudes, thus providing some indication of latitudinal differential rotation. Starspot modeling of the CoRoT time series including differential rotation \cite{Mosser2009b,Ballot2011} returns a rotational frequency at the equator $\Omega_{\rm eq}/2\pi=0.99^{+0.03}_{-0.04}$~$\mu$Hz and $\sin i=0.50^{+0.15}_{-0.16}$ (which again is consistent, see Table 2). The general agreement between all these independent measurements strongly supports our seismic determination of mean internal rotation and inclination angle of HD~52265.

\begin{figure}
\label{fig.rotation}
\centerline{\includegraphics[width=1.0\linewidth]{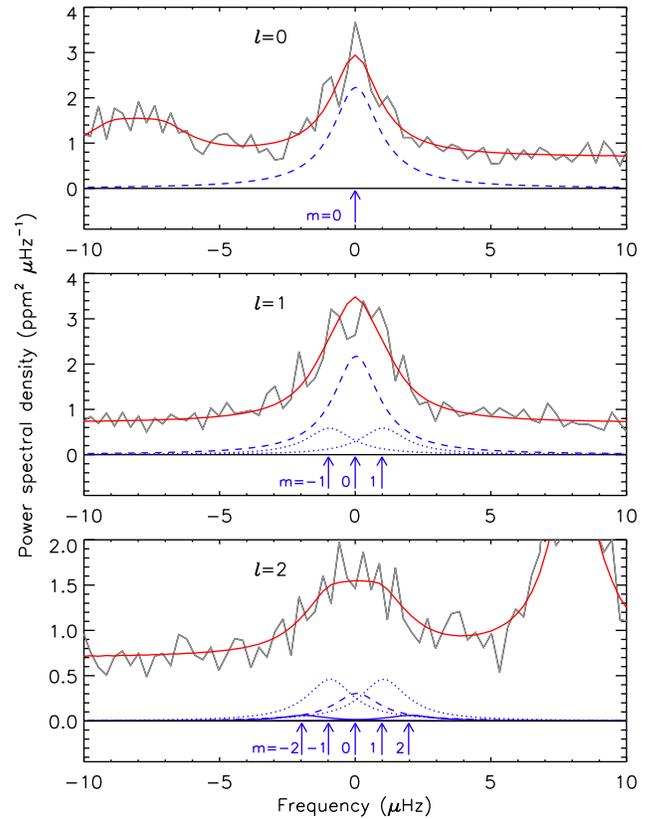}}
\caption{Influence of stellar rotation on oscillations. Power spectra are shown for the radial (top), dipole (middle), and quadrupole modes (bottom), after averaging over the nine consecutive radial orders from Figure~3. The gray curve is the average power spectral density and the red is the average of the fits. For clarity, the frequency resolution is reduced by a factor of 3. Although the rotational splitting is too small to separate the azimuthal components, a rotational broadening of the average line profiles of $l=1$ and $l=2$ is clearly visible. The dashed, dotted, and solid blue lines show the azimuthal components $m=0$, $m=\pm 1$, and $m=\pm 2$, which contribute to the average power. The $2l+1$ frequencies of the azimuthal components, split by rotation ($\Omega/2\pi = 0.98$~$\mu$Hz), are marked by arrows at the bottom of each panel and labelled by the azimuthal order $m$. The visibility amplitudes of the azimuthal components are computed for the best-fit inclination angle of $i=36^\circ$.}
\end{figure}

\begin{figure}
\label{fig.siniomega}
\centerline{\includegraphics[width=\linewidth]{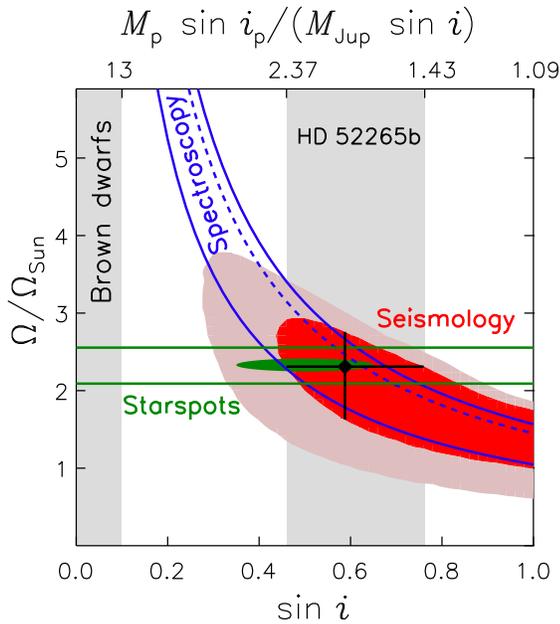}}
\caption{Constraints on stellar rotation and planet mass. The dark-red and light-red regions are the 1-$\sigma$ and 2-$\sigma$ seismic constraints on stellar rotation in the plane $(\Omega/\Omega_{\rm Sun})$\,--\,$(\sin i)$, where $\Omega$ is the bulk angular velocity, $\Omega_{\rm Sun}/2\pi=0.424$~$\mu$Hz is the solar Carrington angular velocity, and $i$ is the inclination of the stellar rotation axis to the line of sight. The black diamond with error bars gives the best-fit seismic values, $\Omega/\Omega_{\rm Sun} = 2.31^{+0.45}_{-0.69}$ and $\sin i = 0.59^{+0.19}_{-0.14}$. For comparison, the two horizontal green lines mark the angular velocity of stellar activity (starspots) deduced from two prominent peaks in the low-frequency part of the power spectrum. The filled green ellipse represents the 1-$\sigma$ bound of the equatorial rotation and inclination angle obtained from starspot modeling of the photometric time series \cite{Mosser2009b,Ballot2011}.
The spectroscopic constraints are given by the dashed (observations) and the solid (1-$\sigma$ errors) blue curves, as expressed through the sky-projected angular velocity $\Omega \sin i = (v \sin i) / R$, where $v \sin i=3.6^{+0.3}_{-1.0}$~\kms is the observed spectroscopic rotational broadening and $R=1.34 R_{\rm Sun}$ is the seismic stellar radius.
The minimum mass of the planet from radial velocity measurements \cite{Butler2006} is $M_{\rm p} \sin i_{\rm p} = (1.09 \pm 0.11) M_{\rm Jupiter}$, where $i_{\rm p}$ is the inclination of the normal of the planetary orbit to the line of sight. Assuming $i_{\rm p}=i$, the seismic constraint on $\sin i$ can be converted into a constraint (top axis and gray region 'HD~52265b') on the true mass of the planet, $M_p$, which is well below the brown dwarf limit of $13 M_{\rm Jupiter}$.
%Starspot angular velocity as a function of latitude, $\lambda$, modelled by $\Omega_{\rm spot}(\lambda)=\Omega_{\rm eq}(1-K\sin^2\lambda)$ with $\Omega_{\rm eq}/2\pi = 0.99^{+0.03}_{-0.04}$~$\mu$Hz and $K=0.25^{+0.05}_{-0.20}$.
}
\end{figure}

\section*{Discussion}
The seismic radius and mass of HD~52265 have been inferred with a formal error of 2\% and the stellar age with a formal error of 5\%\ of the main-sequence lifetime. This level of precision on fundamental stellar properties, required to characterize planets in transit around Sun-like stars \cite{Stello2009,JCD2010,Batalha2011}, is a significant improvement over classical estimates (Table~2). Although the planetary companion orbiting HD~52265 does not transit its host, the seismic determination of the direction of the spin axis of the star provides useful information to further characterize the mass of the exoplanet.

\subsection*{Mass of exoplanet HD~52265b.}
Under the assumption that the rotation axis of the star and the normal to the planetary orbit coincide, i.e., $i_{\rm p}=i$, the knowledge of the minimum mass of HD~52265b from Doppler spectroscopy ($M_{\rm p}\sin i_{\rm p}$) can be used together with the inclination of the stellar spin axis from seismology to constrain the true mass, $M_p$. The 1-$\sigma$ seismic bound $\sin i>0.45$ (irrespective of the value of $\Omega$) implies that $M_{\rm p}<2.4\,{\rm M_{Jupiter}}$. Such a mass is well under the lower limit of brown dwarfs, $M_{\rm BD}= 13 M_{\rm Jupiter}$. Deuterium burning cannot be sustained below this mass threshold. Thus HD\,52265b is likely to be a planet and not 
a brown dwarf.\footnote{Note that the values $\sin i_{\rm p}=0.026$ and $M_{\rm p}=41\,{M_{\rm Jupiter}}$, obtained from Hipparcos intermediate astrometry \cite{Han2001}, have low significance due to insufficient angular resolution of the Hipparcos data \cite{Pourbaix2001b}.} 
The best seismic fit (black cross in Figure~5) implies $1.85^{+0.52}_{-0.42} M_{\rm Jupiter}$. Together, the seismic, spectroscopic, and starspot constraints on $i$ give $M_p = 1.85^{+0.39}_{-0.40} M_{\rm Jupiter}$. Some relevant characteristics of the exoplanet are listed in Table 3.

The formation of Jupiter-mass planets in close orbits (hot Jupiters), like HD 52265b, remains a mystery. In one theory, hot Jupiters are formed in the outer regions of protoplanetary disks and migrate towards the inner regions via friction \cite{Lin1996}. This scenario favors systems with $i=i_{\rm p}$. However, a study based on the Rossiter-McLaughlin effect \cite{Triaud2010} revealed that 80\%\ of a sample of transiting planets show a sky-projected spin-orbit misalignment, $\lambda$, greater than $20^\circ$. Alternative formation scenarios like Kozai cycles \cite{Kozai1962} and planet scattering \cite{Rasio1996,Naoz2011} are required to explain large spin-orbit misalignments. The misalignment of the HD 52265 system is not excluded, as the alignment time scale through tidal interaction \cite{Hut1981} is much greater than the age of the star. For other planets discovered with the transit method, $i_p$ is known and the knowledge of $i$ (through asteroseismology and/or spot modeling) and $\lambda$ would give the true angle between the stellar rotation axis and the planet orbital axis. Asteroseismology has much to contribute to the study of the evolutionary history of exoplanetary systems.

\begin{acknowledgments}
The CoRoT space mission has been developed and is operated by CNES, with the contribution of Austria, Belgium, Brazil, ESA (RSSD and Science Programme), Germany and Spain. L.G. and T.S. acknowledge support from DFG grant SFB 963 ``Astrophysical flow instabilities and turbulence''. J.B. acknowledges support from the ANR project SiROCO. NCAR is supported by the National Science Foundation.

Author contributions: L.G. and T.S. designed research; L.G., J.B., E.M., T.S., G.V., H.B., P.-O.Q., O.B., S.V., T.A., M.A., A.B., C.B., F.B., M.B., T.L.C., C.C., W.J.C., O.L.C., S.D., N.D., Y.E., R.A.G., P.G., S.M., S.M., B.M., C.R., I.W.R., D.S., R.S., K.H.S., and G.A.V. performed research; L.G., J.B., E.M., T.S., S.M.H., and K.R.S. wrote the paper. The authors declare no conflict of interest.
\end{acknowledgments}

\newpage
%\bibliography{literature_hd52265_paper}

\bibliographystyle{pnas}

\end{article}

{
\begin{table*}
\setlength{\arrayrulewidth}{2pt}
\centering
\caption{Parameters of star HD~52265}
\label{table:stellar_parameters}
\vskip10pt
\begin{tabular}{lll}
\hline\\[-10pt]
Distance & $28.95\pm0.34$ pc & astrometry \cite{vanLeeuwen2007} \\
Luminosity & $2.09\pm0.24$ $L_{\rm Sun}$ & astrometry \cite{Ballot2011}\\
Effective temperature, $T_{\rm eff}$ & $6100\pm60$ K & spectroscopy \cite{Ballot2011} \\
Metallicity, [M/H] & $0.19\pm0.05$ & spectroscopy \cite{Ballot2011}\\
Main sequence lifetime & $\approx6 \times 10^9$~yr & mass-luminosity relation \\
\hline
\end{tabular}

\begin{flushleft}
$1 {\rm pc} = 3.086 \times10^{16}$ m; $L_{\rm Sun}=3.8\times10^{26}$ W.
\end{flushleft}

\end{table*}
}

{
\begin{table*}
\setlength{\arrayrulewidth}{2pt}
\centering
\caption{Asteroseismic vs. classical constraints on properties and rotation of HD~52265}
\label{table:seismo_vs_classical}
\vskip10pt
\begin{tabular}{lll}
\hline\\[-5pt]
Stellar property & Asteroseismology & Classical methods\\
\hline\\[-5pt]
Radius ($R_{\rm Sun}$) & $1.34\pm 0.02$ & $1.30\pm0.08$ $\,$ (spectroscopy \cite{Ballot2011}) \\
Mass ($M_{\rm Sun}$)& $1.27\pm 0.03$ & $1.21\pm0.05$ $\,$ (isochrone fits \cite{Valenti2005}) \\[2pt]
Age ($10^9$\,yr) & $2.37 \pm 0.29$ & $2.7^{+0.7}_{-1.5}$ $\,$ (isochrone fits \cite{Valenti2005}) \\
%\hskip0.05in - Bulk rotation, $\Omega/2\pi$ ($\mu$Hz) & $0.98{+0.19\atop-0.29}$ & \\[2pt]
%\hskip0.05in - Starspot rotation, $\Omega_{\rm spot}/2\pi$ ($\mu$Hz) & & $0.91$ and $1.07$ $\quad$ (Photometry \cite{Ballot2011}) \\[2pt]
%\hskip0.05in - Equatorial rotation, $\Omega_{\rm eq}/2\pi$ ($\mu$Hz) & & $0.99{+0.03\atop -0.04}$ $\quad$(Starspot modeling \cite{Ballot2011,Mosser2009b}$\dagger$) \\[2pt]
Bulk rotation, $\Omega/\Omega_{\rm Sun}$ & $2.31^{+0.45}_{-0.69}$ & \\[2pt]
Starspot rotation, $\Omega_{\rm spot}/\Omega_{\rm Sun}$ & & $2.15$ and $2.52$ $\,$ (photometry \cite{Ballot2011}) \\[2pt]
%Equatorial rotation (starspot modeling), $\Omega_{\rm eq}/\Omega_{\rm Sun}$& & $0.99{+0.03\atop -0.04}$ $\quad$(starspot modeling \cite{Ballot2011,Mosser2009b}) \\[2pt]
Inclination of stellar rotation axis, $\sin i$ & $0.59^{+0.18}_{-0.14}$ & $0.50^{+0.15}_{-0.16}$ $\,$ (starspot modeling \cite{Ballot2011,Mosser2009b}) \\[2pt]
Sky-projected rotational velocity (km\,s$^{-1}$) & $3.4 \pm 0.8$ $\dagger$ & $3.6^{+0.3}_{-1.0}$ $\,$ (spectroscopy \cite{Ballot2011})$\ddagger$\\[2pt]
\hline
\end{tabular}

\begin{flushleft}
$\dagger$ $R\Omega\sin i$ where $R$ and $\Omega\sin i$ are seismic estimates.

$\ddagger$ $v \sin i$ from spectroscopic rotational broadening. The value $4.7\pm0.5$ km\,s$^{-1}$ quoted in ref.~\cite{Valenti2005} is probably an overestimate due to the macroturbulence model used.\\

Solar reference values: $R_{\rm Sun}=6.96\times10^8$ m; $M_{\rm Sun}=1.99\times10^{30}$ kg; $\Omega_{\rm Sun}/2\pi=0.424\,\mu$Hz.
\end{flushleft}
\end{table*}
}

{
\begin{table*}
\setlength{\arrayrulewidth}{2pt}
\centering
\caption{Characteristics of exoplanet HD~52265b}
\label{table:planet_parameters}
\vskip10pt
\begin{tabular}{lll}
\hline\\[-10pt]
Minimum planet mass, $M_{\rm p}\sin i_{\rm p}$ & $1.09\pm0.11$ $M_{\rm Jupiter}$ & radial velocity \cite{Butler2006}\\
Semi-major axis of planetary orbit & $0.504 \pm 0.029$ AU & radial velocity \cite{Butler2006}\\[2pt]
Planet mass, $M_{\rm p}$ & $1.85^{+0.52}_{-0.42} M_{\rm Jupiter}\sin i / \sin i_{\rm p}$ & asteroseismology \\[3pt]
\hline
\end{tabular}
\begin{flushleft}
$M_{\rm Jupiter}=1.899\times10^{27}$~kg; $1$~AU$=1.496\times10^{11}$~m.

\end{flushleft}
\end{table*}
}

\end{document}